\documentstyle[12pt]{article}
\def\lsim{\mathrel{\lower2.5pt\vbox{\lineskip=0pt\baselineskip=0pt
           \hbox{$<$}\hbox{$\sim$}}}}
\def\gsim{\mathrel{\lower2.5pt\vbox{\lineskip=0pt\baselineskip=0pt
           \hbox{$>$}\hbox{$\sim$}}}}
\begin{document}
\setlength{\baselineskip}{8mm}
\setcounter{page}{1}
\thispagestyle{empty}
\begin{flushright}
DPNU-95-10 \\
AUE-95/01 \\
April \ 1995 \\
\end{flushright}
\vspace{5mm}
\begin{center}
{\large  \bf
The Weak-Scale Hierarchy \\
and \\
Discrete Symmetries }
\end{center}
\vskip 15mm
\begin{center}
Naoyuki HABA$^1$, Chuichiro HATTORI$^2$,
Masahisa MATSUDA$^3$, \\
Takeo MATSUOKA$^1$  and Daizo MOCHINAGA$^1$ \\
{\it
${}^1$Department of Physics, Nagoya University \\
           Nagoya, JAPAN 464-01 \\
${}^2$Science Division, General Education \\
     Aichi Institute of Technology \\
      Toyota, Aichi, JAPAN 470-03 \\
${}^3$Department of Physics and Astronomy \\
     Aichi University of Education \\
      Kariya, Aichi, JAPAN 448 \\
}
\end{center}
\vspace{10mm}
\begin{abstract}
In the underlying Planck scale theory  we introduce a certain type of
discrete symmetry, which potentially brings the stability of the
weak-scale hierarchy under control.
Under the discrete symmetry the $\mu $-problem and the tadpole problem can be
solved simultaneously without relying on some fine-tuning of parameters.
Instead, it is required that doublet Higgs and color-triplet Higgs fields
reside in different irreducible representations of the gauge symmetry group
at the Planck scale and that they have distinct charges of the
discrete symmetry group.
\end{abstract}
\newpage
Recently, it is greatly expected that many characteristic features of
low-energy effective theory are attributable to various types of symmetry
in the underlying Planck scale theory, such as in superstring theory.
It is plausible that the gauge symmetry $G$ at the Planck scale is larger
than the standard gauge group $G_{st}=SU(3)_C \times SU(2)_L \times U(1)_Y$.
Since the larger $G$ should be broken to $G_{st}$, some $G_{st}$-neutral
fields are needed to be contained in the theory and to develop non-zero
vacuum expectation values (VEVs) at some intermediate energy scales.
And furthermore, it is likely that there exist certain discrete symmetries at
the Planck scale.
As suggested from Gepner model
\cite{Gepner}
, such symmetries may have their origin in symmetrical structure of
compactified space in superstring theory.
The discrete symmetries put some restrictions on interactions including
various couplings related to $G_{st}$-neutral fields.
Then restricted couplings of $G_{st}$-neutral fields to the other fields
reflect on the low-energy effective theory.
In addition, the magnitude of VEVs of $G_{st}$-neutral fields would be
governed by the discrete symmetry and small ratios of the VEVs to the Planck
scale would yield the hierarchical structure to the effective theory.
Consequently, the discrete symmetries and $G_{st}$-neutral fields would
constitute vital ingredients of determining hierarchical structure of the
effective theory.
\par
In constructing realistic unified models we need to treat with a large
hierarchy between two mass scales, i.e. the unification scale and the
electroweak scale
\cite{GUT}.
In general, such models are confronted with the so-called hierarchy problem.
Namely, the weak-scale hierarchy is destabilized by quadratically divergent
radiative corrections.
Supersymmetry (SUSY) is an attractive idea to cure
partially this problem and renders the hierarchy
technically natural
\cite{susy}.
However, the lightness of Higgs doublets
at the tree level is not assured by SUSY and
some fine-tuning of parameters is needed
at the tree level
\cite{fine}.
For example, in the minimal SUSY $SU(5)$ GUT
there appear ${\bf 5}$ and ${\bf 5^*}$ Higgs superfields,
in which Higgs doublets $(H_u, H_d)$ and Higgs triplets
$(g, g^c)$ are contained.
{}From proton stability Higgs triplets should be
superheavy.
While Higgs doublets should not be superheavy
since they constitute an important ingredient of
the low-energy model.
Their mass terms in the superpotential are assumed to be
\begin{equation}
        W \sim \mu H_u H_d + M_g g g^c
\end{equation}
with $\mu =O(10^2$GeV) and $M_g = O(M_{\rm GUT})$.
Why is $\mu $ the electroweak scale but
not the unification scale ?
This is the so-called $\mu $-problem.
Although SUSY protects this mass hierarchy
$M_g \gg \mu $ against radiative corrections,
the mass hierarchy at tree level have to be
fine-tuned.
Many other GUT models also suffer from the $\mu $-problem.
When Higgs doublets and Higgs triplets belong to
the same irreducible representations of $G$,
such as to ${\bf 5}$ and ${\bf 5^*}$ of $SU(5)$,
coupling constants of Higgs doublets and of Higgs triplets to
the singlet or adjoint Higgs are of the same order.
In order to get triplet-doublet mass splitting
without fine-tuning,
we are enforced to introduce additional Higgs fields with
the larger representations of $G$.
However, it seems that these enlargements of
the Higgs sector are rather complicated and
bring about another problem to the models
\cite{missing}.
Therefore, it is likely that
Higgs doublets and Higgs triplets reside in
different irreducible representations of
the gauge group $G$ at the Planck scale.
\par
In anticipation of explaining $\mu = O(10^2$GeV),
several authors introduced a $G_{st}$-neutral field $N$
\cite{singlet},
provided that Higgs triplets have a large mass $M_g$
whereas Higgs doublets remain massless at
the unification scale.
In this scenario there exist trilinear couplings
with $N$.
The superpotential is given by
\begin{equation}
       W \sim M_g g g^c + f_H N H_u H_d + f_g N g g^c \ .
\end{equation}
Unless we have some kinds of selection rule
on trilinear couplings,
the coupling constants $f_H$ and $f_g$ are to be $O(1)$.
Suppose $N$ develops a nonzero VEV with $O(10^2$GeV)
via some mechanism,
the $\mu $-term is induced as
\begin{equation}
      \mu = f_H \langle N \rangle = O(10^2{\rm GeV}).
\end{equation}
Even if this is the case, however,
we encounter a new hierarchical problem.
A trilinear coupling of the singlet $N$ to superheavy
Higgs triplets $g, g^c$ brings about large mass
correction to Higgs doublet scalar fields through
tadpole diagrams as shown in Fig.1.
\vskip 0.5cm
\begin{center}
 \unitlength=0.8cm
 \begin{picture}(2.5,2.5)
  \thicklines
  \put(0,0){\framebox(3,1){\bf Fig. 1}}
 \end{picture}
\end{center}
\vskip 0.5cm
The contribution of tadpole diagrams to Higgs scalar mass
is given by
\begin{equation}
   \delta m_{H_u,H_d}^2 \sim
                 \frac {M_g f_g f_H m_{3/2}^3}{m_N^2},
\end{equation}
where $m_N$ represents the scalar mass of $N$.
Since the soft SUSY breaking terms give the scalar mass,
$m_N$ becomes of the order of $m_{3/2} = O(1$TeV).
This mass correction is extremely large compared to
$O(m_{3/2}^2)$.
Thus the coupling of $N$ to $g, g^c$ destabilizes
the mass hierarchy $M_g \gg \mu $.
This is the so-called tadpole problem or
light singlet problem
\cite{tadpole}.
\par
In this paper we propose a new model with
a certain type of discrete symmtery.
The discrete symmetry implies a stringent selection rule
on renormalizable and nonrenormalizable interactions
given by the superpotential.
In the model a mirror pair of $G_{st}$-neutral
fields $N$ and ${\overline N}$ is contained and
develops a very large VEV $\langle N \rangle =
\langle {\overline N} \rangle $.
Without relying on some fine-tuning among parameters
we obtain the relations
\begin{equation}
        \mu = f_H \langle N \rangle , \ \ \ \ \
        f_H < O\left( \frac {m_{3/2}}
                 {\langle N \rangle } \right), \ \ \ \ \
        f_g = O(1)
\end{equation}
for effective couplings, respectively,
with $M_g = O(\langle N \rangle ) \gg m_{3/2}$ and
$m_N^2 = O(m_{3/2}^2)$.
The smallness of the coupling $f_H$ is explained naturally
from the discrete symmetry.
It follows that
\begin{equation}
        \mu < O(m_{3/2}), \ \ \ \ \
        \delta m_{H_u,H_d}^2 = O(\mu \,m_{3/2}).
\end{equation}
In this model the $\mu $-problem and the
tadpole problem are closely linked together and
solved simultaneously.
\par
In the model proposed here we are based on the following
scheme of superstring or supergravity models.
The gauge symmetry group $G$ at the Planck scale
$\Lambda _P$ is rank-five or rank-six, such as
in $E_6$-inspired models.
In matter superfields there appear doublet Higgs superfields
$H_u, H_d$ and color-triplet Higgs superfields $g, g^c$ which
reside in distinct irreducible representations of $G$.
In addition to these chiral superfields, we have
a mirror pair of $G_{st}$-neutral but $G$-charged chiral
superfields $N$ and ${\overline N}$.
Existence of mirror fields is likely in superstring models.
It is supposed that as far as gauge invariance is concerned,
the couplings $N H_u H_d$ and $N g g^c$ are allowed
whereas the couplings ${\overline N} H_u H_d$ and
${\overline N} g g^c$ are forbidden.
\par
Let us introduce certain discrete symmetries at
the Planck scale,
which may be a reflection of the geometrical structure
of the compactified space.
In fact, peculiar discrete symmetries come into Gepner
model in which the compactified space is constructed
algebraically by a tensor product of $N = 2$
superconformal field theory
\cite{Gepner}.
Concretely, the discrete symmetry $Z_{k+2}$ or
$Z_{k+2} \times Z_2$ is derived from $N = 2$ superconformal
field theory with the level $k$
in which each matter superfield has a distinct charge
of the discrete symmetries.
The discrete symmetries put a stringent selection rule
on allowed couplings in the superpotential.
In the present model it is assumed that allowed couplings
are given by
\begin{equation}
      W = \frac {\lambda _H}{\Lambda _P^{2p}} \,
                       (N {\overline N})^p N H_u H_d \
           + \ \lambda _g N g g^c \
           + \ \frac {\lambda _N}{\Lambda _P^{2l-1}} \,
                       (N {\overline N})^{l+1}
           + \cdots ,
\end{equation}
where $1 \leq p, l$ and the coefficients $\lambda _H,
\lambda _g, \lambda _N $ are $O(1)$.
As we will see later, the exponents $p$ and $l$ are determined
according as the discrete charges of the matter superfields.
Our assumption contains that there appears a trilinear
coupling only for colored Higgs fields
because of special values of the discrete charge of
the products $(H_uH_d)$ and $(gg^c)$.
\par
Incorporating the soft SUSY breaking terms, we can get
the scalar potential $V$.
The scale of SUSY breaking $(m_{3/2})$ is supposed to be
$O(1$TeV).
The running scalar masses squared $m_N^2$ and
$m_{\overline N}^2$ for $N$ and ${\overline N}$ are
$O(m_{3/2}^2)$.
Since $N$ couples to colored Higgs with a sizable trilinear
coupling constant,
$m_N^2$ possibly becomes negative even at
large energy scale
\cite{Zoglin}.
When $m_N^2 + m_{\overline N}^2 < 0$,
$N$ and ${\overline N}$ develop nonzero VEVs.
By minimizing $V$,
we obtain the VEVs
\cite{discrete}
\cite{majom}
\begin{equation}
    \langle N \rangle = \langle {\overline N}\rangle
         \equiv \Lambda \sim \Lambda _P
           \left( \frac {m_{3/2}}{\Lambda _P}\right) ^{1/{2l}},
\end{equation}
which is sufficiently large compared to $m_{3/2}$.
For instance, we have $\Lambda \gsim 10^{16}$GeV
for $l \geq 3$.
The magnitude of the scale $\Lambda $ is controlled by
the discrete charges of $N$ and $\overline N$.
Although spontaneous breaking of the gauge symmetry
occurs at the scale $\Lambda $,
the $D$-flatness condition is satisfied and then
SUSY is preserved at this scale.
In the symmetry breaking a combination
$(N-{\overline N})/{\sqrt 2}$ is absorbed by a vector
superfield due to the Higgs mechanism.
The remaining component $(N + {\overline N} - 2\Lambda )
/{\sqrt 2} \ (\equiv N')$ has a mass of order $O(m_{3/2})$
irrespective of $l$.
For the sake of convenience we introduce
the notation $x$ defined by
\begin{equation}
     x = \frac {\Lambda }{\Lambda _P}
             \sim \left( \frac {m_{3/2}}
                {\Lambda _P}\right) ^{1/{2l}}.
\end{equation}
This small ratio $x$ becomes an efficient parameter
in describing the hierarchical structure of the effective
theory.
\par
Now we proceed to study the low-energy effective
superpotential $W^{eff}$ below the scale $\Lambda $.
{}From Eqs. (7) to (9) the bilinear terms in $W^{eff}$
becomes
\begin{equation}
      W_2^{eff} = \mu \,H_u H_d + M_g \,g g^c + M_N \,N'^2
\end{equation}
with
\begin{eqnarray}
     \mu & \simeq & \lambda _H \,x^{2p+1}\Lambda _P, \\
     M_g & = &      \lambda _g \,\Lambda
                       = \lambda _g \, x \Lambda _P, \\
     M_N & \simeq & \lambda _N \,x^{2l}\Lambda _P
                       \simeq m_{3/2}.
\end{eqnarray}
Colored Higgs fields get a mass of $O(\langle N \rangle )$,
while doublet Higgs mass $\mu $ is controlled
by the exponent $p$.
Explicitly, $\mu $ is given by
\begin{equation}
      \mu \simeq x^{2(p-l)+1} \,m_{3/2}.
\end{equation}
Therefore, when $p \geq l$,
the $\mu $-problem is solved.
In what follows we take the condition
\begin{equation}
            p \geq l
\end{equation}
and then $\mu \lsim x \,m_{3/2} < O(1$TeV).
For example, we obtain $\mu = O(10^2$GeV) for $p = l \sim 8$.
\par
To address ourselves to the tadpole problem,
we study the trilinear terms in $W^{eff}$
which are of the form
\begin{equation}
       W_3^{eff} = f_H \,N' H_u H_d
                   + f_g \,N' g g^c + f_N\, N'^3,
\end{equation}
where
\begin{equation}
      f_H \simeq \lambda _H \,x^{2p}, \ \ \ \ \
      f_g = \lambda _g/{\sqrt 2}, \ \ \ \ \
      f_N \simeq \lambda _N \,x^{2l-1}.
\end{equation}
As a consequence of the discrete symmetry
it follows that we have
\begin{equation}
        f_g = O(1)
\end{equation}
for $N' g g^c$ coupling, whereas
\begin{equation}
        f_H = \frac {\mu }{\Lambda } \ll 1
\end{equation}
for $N' H_u H_d$ coupling.
The tadpole contribution to
the Higgs mass becomes
\begin{equation}
     \delta m_{H_u,H_d}^2 \simeq
             \frac {M_g f_g f_H m_{3/2}^3}{m_N^2}
             \simeq M_g f_H m_{3/2}
             \simeq \mu \,m_{3/2}.
\end{equation}
This implies that the tadpole problem is solved
simultaneously together with the $\mu $-problem
under the condition (15).
This is due to the fact that both the $\mu $-term
and the trilinear coupling $N'H_uH_d$ are induced from
the nonrenormalizable interaction
$(N{\overline N})^p\,N H_u H_d$ in the underlying theory.
\par
For illustration we take up $Z_{\alpha }$ as a simple
example of the discrete symmetry,
where $\alpha $ is an integer larger than one.
As mentioned above, this type of discrete symmetry
possibly comes into Calabi-Yau string models.
In Table I we tabulate $Z_{\alpha }$-charges of
matter superfields,
where $b$ and $c$ represent $Z_{\alpha }$-charges of
the products $(H_u H_d)$ and $(g g^c)$, respectively.
Generally, as is the case with Gepner model,
Grassmann number $\theta $ also has
a nonzero charge denoted as $-d$ in Table I.
Each charge is taken as $0  \leq a, {\overline a},
b,c, d < \alpha $.
Since the superpotential (7) is assumed to be
a consequence of the $Z_{\alpha }$ symmetry,
we have the relations
\begin{equation}
\left.
\begin{array}{r}
         p (a + {\overline a}) + a + b + 2d  \equiv 0 \\
         a + c + 2d  \equiv  0  \\
         (l+1)(a + {\overline a}) + 2d  \equiv  0
\end{array}
\right\}
\ \ \ {\rm mod}\ \alpha ,
\end{equation}
where $1 \leq p, \, l < \alpha $.
When $a, {\overline a}, b, d$ are given,
$c$ and the exponents $p$ and $l$ are determined from these
equations.
To get a nontrivial solution, the conditions
$ a+{\overline a},\ b-c \not \equiv 0 $ (mod $\alpha $)
should be satisfied.
The above relations lead to
\begin{equation}
      (p-l)(a+{\overline a}) + b-{\overline a} \equiv 0
                  \ \ \ {\rm mod } \ \alpha .
\end{equation}
Thus, if $b \equiv {\overline a}$ and if
$a+{\overline a}$ is prime to $\alpha $,
we obtain
\begin{equation}
       p=l.
\end{equation}
This case is in accord with the condition (15).
More concretely, when $a={\overline a}=b=d=1$ and
$\alpha $ is odd, we get $p=l=c+1=\alpha -2$.
\vskip 0.5cm
\begin{center}
 \unitlength=0.8cm
 \begin{picture}(2.5,2.5)
  \thicklines
  \put(0,0){\framebox(3,1){\bf Table I}}
 \end{picture}
\end{center}
\vskip 0.5cm
\par
As for the generation structure of the $G_{st}$-neutral
fields,
so far it is postulated that we have
only a pair of $N$ and ${\overline N}$.
Generally, however, the multiplicities
of $N$ and of the mirror superfield ${\overline N}$
do not coincide with each other
but rather in superstring models
the difference of these multiplicities
corresponds to the generation number.
Taking this situation into consideration,
we change the above model with a pair of $N$ and
${\overline N}$ for another model with a double
$G_{st}$-neutral field $N_0, N_1$ and a single mirror
field ${\overline N}$.
In this case the discrete symmetry is put to
$Z_{\alpha } \times Z_2$ (or $Z_{2\alpha }$).
As suggested from superstring models,
matter fields would have individual discrete
charges for every generation.
If $N_0$, ${\overline N}$, $(H_uH_d)$ and $(gg^c)$
are all even under $Z_2$ and if $N_1$ is odd,
the superpotential $W$ does not contain odd terms
with respect to $N_1$ due to the $Z_2$ symmetry.
Indeed, the superpotential is written as
\begin{eqnarray}
      W & = & \frac {\lambda _H}{\Lambda _P^{2p}} \,
                       (N_0 {\overline N})^p N_0 H_u H_d \
           + \ \lambda _g N_0 g g^c \
           + \ \frac {\lambda _N^{(0)}}{\Lambda _P^{2l-1}} \,
                       (N_0 {\overline N})^{l+1} \nonumber \\
        & & \makebox[5em]{}  + \ \frac {\lambda _N^{(2)}}
                                         {\Lambda _P^{2n-1}} \,
              (N_0 {\overline N})^{n-1} (N_1 {\overline N})^2 \,
                + \cdots .
\end{eqnarray}
Since $N_0$ has a sizable trilinear coupling to $gg^c$
while $N_1$ does not,
it is natural that the running scalar
mass squared $m_{N_0}^2$ becomes negative but
$m_{N_1}^2$ remains positive at large energy scale.
When $m_{N_0}^2 + m_{\overline N}^2 < 0$,
$N_0$ and ${\overline N}$ develop nonzero VEVs
whereas $\langle N_1 \rangle = 0$.
In view of the circumstances it follows that
the present model exhibits hierarchical structure
of the effective superpotential
quite similar to that of the previous model.
Therefore, under the condition $p \geq l$ the $\mu $-problem
and the tadpole problem are solved also in this model.
It is expected that the considerations described here
can be reasonably generalized to the models
with more complicated generation structure of matter fields.
\par
In conclusion, a certain type of discrete symmetries
for the underlying Planck scale theory can control
the weak-scale hierarchy of the effective theory.
Under the discrete symmetries the $\mu $-problem
and the tadpole problem are closely linked together
and are solved simultaneously.
This is because the $\mu $-term and the trilinear coupling
$NH_uH_d$ in the low-energy effective theory have
their origins in the common nonrenormalizable interaction.
The solution is assured by the condition (15).
It should be emphasized that we do not rely on
some fine-tuning of parameters.
Instead, it is required that doublet Higgs and
colored Higgs fields reside in different irreducible
representations of the gauge symmetry at the Planck
scale and that they have distinct charges of the
discrete symmetry.
In view of the phenomenological result that
in the minimal supersymmetric standard model
gauge couplings are unified at the scale
$O(10^{16}$GeV) smaller than the Planck scale
\cite{MSSM},
it is tempting to find GUT-type models consistent
with such particle assignments.
As pointed out by the authors
\cite{aligned},
there are such GUT-type string models.
It is very interesting to construct phenomenologically
viable GUT-type models which satisfies the condition (15).

\newpage
{\bf\large Figure Captions}
\vskip 1cm
{\bf Fig. 1}\ \ \ \ \
A tadpole diagram which contributes to Higgs scalar mass.
$H_u$, $H_d$ and $g$, $g^c$ stand for doublet Higgs and
colored Higgs fields, respectively.
The $G_{st}$-neutral field is denoted as $N$.
\vskip 4cm
{\bf\large Table Captions}
\vskip 1cm
{\bf Table I}\ \ \ \ \
Charges of the discrete symmetry $Z_{\alpha }$ for
matter superfields.
$b$ and $c$ represent charges of the products
$(H_uH_d)$ and $(gg^c)$, respectively.
In general, Grassmann number $\theta $ also has
a nonzero charge denoted as $-d$.
\vskip 1cm
\begin{center}
{\bf Table I }\\
\vspace {1cm}
\begin{tabular}{|c|c|}  \hline
       Fields     &   $Z_{\alpha }$-charges      \\ \hline
        $N$       &        $a$                   \\
 ${\overline N}$  &     ${\overline a}$          \\
    $(H_uH_d)$    &        $b$                   \\
     $(gg^c)$     &        $c$                   \\
    $\theta $     &       $-d$                   \\ \hline
\end{tabular}
\end{center}

\newpage
\begin{thebibliography}{1}
\bibitem{Gepner}
D. Gepner, Phys. Lett. {\bf 199B} (1987) 380;
           Nucl. Phys. {\bf B296} (1988) 757.
\bibitem{GUT}
H. Georgi and S. L. Glashow, Phys. Rev. Lett. {\bf 32} (1974) 438.
\bibitem{susy}
M. Veltman, Acta Phys. Pol. {\bf B12} (1981) 437. \\
S. Dimopoulos and S. Raby, Nucl. Phys. {\bf B192} (1981) 353. \\
E. Witten, Nucl. Phys. {\bf B188} (1981) 513. \\
M. Dine, W. Fischler and M. Srednicki, Nucl. Phys. {\bf B189}
                           (1981) 575.
\bibitem{fine}
E. Witten, Nucl. Phys. {\bf B185} (1981) 513. \\
S. Dimopoulos and H. Georgi, Nucl. Phys. {\bf B193} (1981) 150. \\
N. Sakai, Z. Phys. {\bf C11} (1981) 153.
\bibitem{missing}
H. Georgi, Phys. Lett. {\bf 108B} (1982) 283. \\
B. Grinstein, Nucl. Phys. {\bf B206} (1982) 387. \\
R. N. Cahn, I. Hinchliffe and L. J. Hall, Phys. Lett.
                   {\bf 109B} (1982) 426. \\
A. Masiero, D. V. Nanopoulos, K.Tamvakis and T. Yanagida,
                   Phys. Lett. {\bf 115B} (1982) 380. \\
A. Buras, J. Ellis, M. Gaillard and D. V. Nanopoulos,
                   Nucl. Phys. {\bf B135} (1985) 66. \\
K. S. Babu and S. M. Barr, Phys. Rev. {\bf D48} (1993) 5354. \\
J. Hisano, H. Murayama and T. Yanagida, Phys. Rev. {\bf D49}
                             (1994) 4966.
\bibitem{singlet}
E. Witten, Phys. Lett. {\bf 105B} (1981) 267. \\
L. Ibanez and G. G. Ross, Phys. Lett. {\bf 110B} (1982) 215. \\
D. V. Nanopoulos and K. Tamvakis, Phys. Lett. {\bf 113B}
                     (1982) 151. \\
S. Dimopoulos and H. Georgi, Phys. Lett. {\bf 117B} (1982) 287.
\bibitem{tadpole}
J. Polchinski and L.Susskind, Phys. Rev. {\bf D26}
                                        (1982) 3661. \\
H. P. Nills, M. Srednicki and D. Wyler, Phys. Lett.
                                {\bf 124B} (1983) 337. \\
A. B. Lahanas, Phys. Lett. {\bf 124B} (1983) 341. \\
J. Bagger and E. Poppitz, Phys. Rev. Lett. {\bf 71}
                                       (1993) 2380.
\bibitem{Zoglin}
P. Zoglin, Phys. Lett. {\bf 228B} (1989) 47.
\bibitem{discrete}
C. A. L\"utkin and G. G. Ross, Phys. Lett. {\bf 214B}
                                     (1988) 357. \\
C. Hattori, M. Matsuda, T. Matsuoka and H. Mino,
             Prog. Theor. Phys. {\bf 82} (1989) 599.
\bibitem{majom}
N. Haba, C. Hattori, M. Matsuda T. Matsuoka
          and D. Mochinaga, Phys. Lett. {\bf B337} (1994) 63;
             Prog. Theor. Phys. {\bf 92} (1994) 153.
\bibitem{MSSM}
U. Amaldi, W. de Boer and H. Furstenau, Phys. Lett. {\bf 260B}
                (1991) 447. \\
P. Langacker and M. Luo, Phys. Rev. {\bf D44} (1991) 817. \\
J. Ellis, S. Kelly and D. V. Nanopoulos, Phys. Lett.
                   {\bf 249B} (1990) 441.
\bibitem{aligned}
N. Haba, C. Hattori, M. Matsuda T. Matsuoka
             and D. Mochinaga, preprint DPNU-94-37
                     AUE-06-94 (1994) hep-ph/9409361.
\end{thebibliography}
\end{document}